\documentclass[journal]{IEEEtran}
\IEEEoverridecommandlockouts

\usepackage{amsmath,amssymb}
\usepackage{graphicx}
\usepackage{booktabs}
\usepackage{multirow}
\usepackage{xcolor}
\usepackage{tikz}
\usetikzlibrary{arrows.meta,positioning,fit,calc}
\usepackage[hidelinks]{hyperref}

\begin{document}

\title{Self-Supervised Deconvolution of In-Air Sonar Images Using Sensor Ego-Motion}

\author{Jan~Steckel%
\thanks{J. Steckel is with Cosys-Lab, University of Antwerp, 2020 Antwerp, Belgium, and with Flanders Make Strategic Research Centre, Lommel, Belgium (e-mail: jan.steckel@uantwerpen.be).}%
}

\markboth{IEEE Signal Processing Letters}{Steckel: Self-Supervised Deconvolution of In-Air Sonar Images Using Sensor Ego-Motion}

\maketitle

\begin{abstract}
Broadband in-air sonar sensors with a sparse microphone array form images through delay-and-sum beamforming, which leaves high sidelobes and an angular resolution that degrades strongly towards the sides of the field of view. Deconvolving these images requires access to the point-spread function of the complete imaging chain, including the emitter spectrum and directivity and the array imperfections. In practise, these aspects are hard to calibrate for. In this letter, we propose a self-supervised method that learns a deconvolution network together with a physics-structured model of the point-spread function, requiring only the sonar data itself and the odometry of a moving robot on which the sensor is mounted. Using acoustic flow warping, the internal representation of the model is warped and subsequently rendered through the learned point-spread function. This is then compared to the measured images under a speckle likelihood in accordance with the statistical model of image formation. In simulation, the learned model recovers the unknown emitter directivity, and the deconvolved images reach a constant angular resolution of 5$^\circ$ to 6$^\circ$ over the full field of view, against 12$^\circ$ to 30$^\circ$ for delay-and-sum and 7$^\circ$ to 23$^\circ$ for delay-multiply-and-sum with coherence factor. The method outperforms classical deconvolution with the nominal point-spread function, and even outperforms CLEAN with the true one.
\end{abstract}


\section{Introduction}
\IEEEPARstart{I}{n-air} sonar is a mature sensing modality for mobile robots in conditions where cameras and lidar sensors fail, such as the presence of dust, fog or darkness. Indeed, 3D sonar sensors with a sparse array of MEMS microphones and a single broadband emitter have been used for navigation \cite{steckelBroadband3DSonar2013, steckelAcousticFlowBasedControl2017, jansenRealTimeSonarFusion2022}, mapping \cite{steckelBatSLAMSimultaneousLocalization2013} and in harsh field conditions \cite{kerstensERTISFullyEmbedded2019, laurijssenRuggedizedEmbeddedUltrasound2026}. Their images (so-called energyscapes) are formed by matched filtering, Delay-And-Sum (DAS) beamforming and envelope detection \cite{jansenInAirImagingSonar2019}. With 32 microphones distributed pseudo-randomly in an aperture of a few centimeters, the resulting Point-Spread Function (PSF) has high sidelobes, and its main lobe widens strongly towards the sides of the field of view. Larger arrays \cite{ laurijssenHiRISAirborneSonar2024} come at a cost in size, power and price, while regular arrays \cite{verellenHighResolutionUltrasoundSensing2020} only work well in narrowband formulations.

Recently, we showed that Delay-Multiply-And-Sum (DMAS) beamforming with Coherence Factor (CF) weighting \cite{matroneDelayMultiplySum2015, mallartAdaptiveFocusingScattering1994} strongly increases the dynamic range of in-air acoustic images in real time \cite{jansenDelayMultiplyandSumBeamformingRealTime2026}, but DMAS does not remove the widening of the main lobe. Deconvolution methods invert the PSF instead, such as DAMAS \cite{brooksDeconvolutionApproachMapping2006, doughertyExtensionsDAMASBenefits2005}, CLEAN-SC \cite{sijtsmaCLEANBasedSpatial2007} or Richardson-Lucy \cite{richardsonBayesianBasedIterativeMethod1972, lucyIterativeTechniqueRectification1974}, and neural networks perform this deconvolution much faster \cite{maPhasedMicrophoneArray2019, kujawskiDeepLearningMethod2019, xuAcousticSourceImaging2021, pintoDeconvolutingAcousticBeamforming2021, liangLearningInterpretableEndtoend2024}, and learned beamformers have been proposed for medical ultrasound imaging \cite{luijtenAdaptiveUltrasoundBeamforming2020}. All of these need an accurate PSF or ground-truth source maps, typically from a model of the array, or accurate ground-truth measurements of the scene layout. For an active sonar, however, the PSF also contains the emitter spectrum and directivity and the mismatch of the microphones. Self-calibration \cite{friedlanderEigenstructureMethodsDirection1988} and blind deconvolution \cite{renNeuralBlindDeconvolution2020} estimate parts of this from data, but from a single image, the PSF and the scene are fundamentally ambiguous.

Mounting the sensor on a moving robot can resolve this ambiguity. When the sensor moves, a reflector moves through the image according to the acoustic flow equations \cite{steckelAcousticFlowBasedControl2017, jansenAdaptiveAcousticFlowBased2021}, which only depends on the ego-motion and on the position of the reflector in the scene. A sidelobe does not follow its own acoustic flow, but follows the reflector that caused the sidelobe to exist. Predicting future measurements has been used before as a learning signal for in-air sonar \cite{steckelEchoPTPretrainedTransformer2024, jansenSonoNERFsNeuralRadiance2024}, and invariances of the measurement process allow learning to image without ground truth \cite{chenEquivariantImagingLearning2021, lehtinenNoise2NoiseLearningImage2018}. In this letter, a network maps a DAS energyscape to a sparse reflectivity map, which is warped to the poses of the next measurements using the robot's odometry and rendered through a learned, physics-structured model of the imaging chain. Both the deconvolution and the PSF of the sensor are thereby learned from sonar data and odometry only. We validate the approach in simulation against DAS, DMAS and classical deconvolution approaches.

\begin{figure*}[t]
\centering
\includegraphics[width=\linewidth]{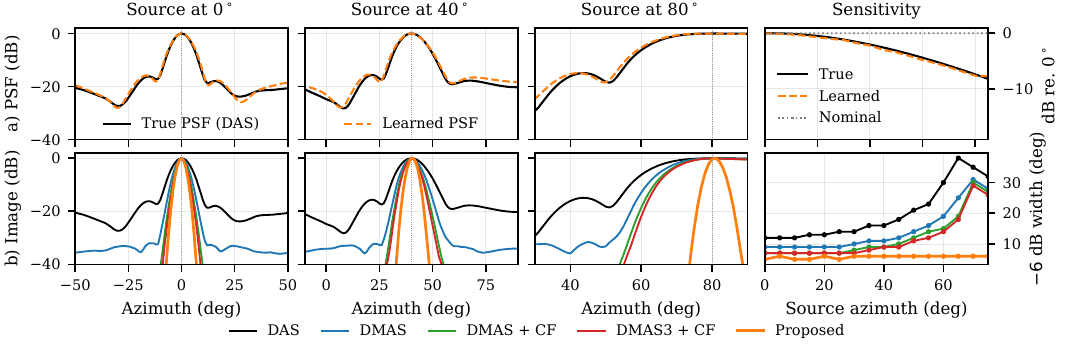}
\caption{Point-spread functions for a point reflector at 2~m and source azimuths of 0$^\circ$, 40$^\circ$ and 80$^\circ$ (maximum over $\pm 1$ range cell). a) True PSF of the simulated sensor (DAS) and the PSF learned without ground truth; right: sensitivity over the field of view. b) Output of the complete imaging chain; right: $-6$~dB width versus source azimuth. The deconvolved images keep a width of 5$^\circ$ in all directions, whereas DAS and DMAS widen by a factor 3.}
\label{fig:psf}
\end{figure*}

\section{Imaging Model}
\label{sec:model}
We consider an eRTIS-type sensor \cite{kerstensERTISFullyEmbedded2019, laurijssenRuggedizedEmbeddedUltrasound2026} with one emitter and $N=32$ microphones in the $YZ$-plane, at positions $\begin{bmatrix} 0 & p_y(i) & p_z(i) \end{bmatrix}^T$. The emitter sends a broadband FM sweep, and each microphone signal $s_{M_i}[k]$ is first processed by a matched filter \cite{steckelBroadband3DSonar2013}:
\begin{equation}
s_{MF_i}[k] = \mathcal{F}^{-1}\left\{ S_{M_i}(j\omega) \cdot S_b^*(j\omega) \right\}
\end{equation}
with $S_{M_i}(j\omega)$ the DFT of the $i$-th microphone signal and $S_b^*(j\omega)$ the complex conjugate of the DFT of the emitted signal. Next, a beamformer steers the array into each direction of interest $\psi$, with azimuth $\theta$ and elevation $\varphi$:
\begin{equation}
s^{BF}_{\psi}[k] = \frac{1}{N}\sum_{i=1}^{N} s_{MF_i}\left[k + \tau_i(\psi)\right]
\label{eq:das}
\end{equation}
where the delay $\tau_i(\psi) = f_s\cdot(\psi_y(i) + \psi_z(i))/c$, with $\psi_y(i) = \sin(\theta)\cos(\varphi)\,\bar{p}_y(i)$ and $\psi_z(i) = \sin(\varphi)\,\bar{p}_z(i)$, is calculated from the nominal microphone positions $\bar{p}_y(i)$, $\bar{p}_z(i)$, with $f_s$ the sample rate and $c$ the speed of sound. The envelope $s^{EN}_{\psi}[k]$ of each beamformed signal, averaged in power over range cells of width $\Delta r$, forms the energyscape $P(r,\psi)$ \cite{steckelBroadband3DSonar2013}. We restrict ourselves to 2D energyscapes in this letter ($\varphi = 0$) with $M$ directions of interest, and denote the energyscape of pulse $t$ as $P_t(r,\psi)$.

Following \cite{steckelBroadband3DSonar2013}, we obtain the Point-Spread Function (PSF) of this imaging chain through a data generation model. A unit point reflector in direction $\psi'$ produces the matched-filtered signal $\mathcal{F}^{-1}\{ g_i\, S(j\omega)\, G(\psi',\omega)\, e^{-j\omega \Delta_i(\psi')}\}$ at microphone $i$, with $g_i$ the complex gain of the microphone, $S(j\omega)$ the spectrum of the emitted and matched-filtered pulse, $G(\psi',\omega)$ the directivity of the sensor, and $\Delta_i(\psi') = \big(\tilde{\psi}'_y(i) + \tilde{\psi}'_z(i) + \vec{e}^{\,T}\vec{u}(\psi')\big)/c$ its delay. Here, $\tilde{\psi}'_y(i)$ and $\tilde{\psi}'_z(i)$ are calculated as above, but from the true microphone positions $p_y(i)$, $p_z(i)$, $\vec{e}$ is the position of the emitter relative to the array center, and $\vec{u}(\psi')$ the unit vector in direction $\psi'$. Beamforming these signals with equation \ref{eq:das} (nominal positions), taking the envelope and the energy in range cell $\delta$ relative to the reflector gives the PSF $K(\delta,\psi,\psi')$. Neglecting the interference between reflectors in the same range cell, the expected energyscape of a reflectivity map $E(r,\psi')\geq 0$ is then given by:
\begin{equation}
\hat{P}(r,\psi) = \sum_{\psi'}\sum_{\delta} K(\delta,\psi,\psi')\, a(r_\delta)\, E(r_\delta,\psi')
\label{eq:render}
\end{equation}
with $r_\delta = r - \delta\,\Delta r$, and $a(r) = a_0\, r^{-4}\, 10^{-2\alpha r/10}$ accounting for spherical spreading and air absorption $\alpha$ (dB/m). In practice, the model parameters ($g_i$, $S$, $G$, $p_y(i)$, $p_z(i)$, $\vec{e}$ and $\alpha$) are not accurately known but do impact image formation. We therefore make all of these parameters learnable, with $G$ represented by a small multilayer perceptron on Fourier features of $\vec{u}(\psi')$ with one output in dB per frequency, and denote the resulting PSF as $K_\Theta$, with $\Theta$ the set of model parameters. We initialize them with the nominal design: the nominal array, a flat directivity, the spectrum of the call and the nominal emitter position. As only the product of $a_0$ and $E$ enters equation \ref{eq:render}, the gain $a_0$ cannot be learned directly. Therefore, we fix this value from the data preliminarily, so that a reflector with $E=1$ produces the strongest echoes observed in the energyscapes.

\section{Self-Supervised Deconvolution}
\label{sec:method}
\begin{figure*}[t]
\centering
\resizebox{0.95\textwidth}{!}{%
\begin{tikzpicture}[
  node distance=4mm and 6mm,
  blk/.style={draw, rounded corners=1.5pt, align=center, font=\footnotesize, minimum height=9mm, inner sep=2pt},
  dat/.style={blk, fill=gray!12},
  lrn/.style={blk, fill=orange!18},
  fix/.style={blk, fill=blue!10},
  sub/.style={draw, rounded corners=1pt, fill=orange!18, font=\scriptsize, minimum height=3.8mm, minimum width=27mm, inner sep=1pt},
  arr/.style={-{Stealth[length=2mm]}, thick},
]
\node[dat] (Pt) {DAS energyscape\\ $P_t(r,\psi)$};
\node[sub, right=11mm of Pt, yshift=6.5mm] (b1) {Dense mixing $M{\times}M$};
\node[sub, below=0.8mm of b1] (b2) {Graph conv.\ (Chebyshev)};
\node[sub, below=0.8mm of b2] (b3) {Range conv.\ (dilated)};
\node[draw, dashed, rounded corners=1pt, inner sep=1.5pt, fit=(b1)(b2)(b3)] (rb) {};
\node[font=\scriptsize, anchor=west, inner sep=1pt] at (rb.east) {$\times 6$};
\node[sub, below=2.5mm of rb] (sp) {Softplus $\rightarrow \tilde{E}_t \geq 0$};
\draw[-{Stealth[length=1.3mm]}] (rb) -- (sp);
\node[draw, rounded corners=2pt, inner xsep=5mm, inner ysep=2pt, fit=(rb)(sp)] (grp) {};
\node[font=\scriptsize, anchor=south] at (grp.north) {Deconvolution network $f_\phi$};
\node[dat, right=8mm of grp] (E) {Clean map\\ $E_t = H * \tilde{E}_t$};
\node[fix, right=10mm of E] (warp) {Acoustic-flow\\ warp $\mathcal{W}_k$, (\ref{eq:flow})};
\node[lrn, right=12mm of warp] (fwd) {Forward model $K_\Theta$, (\ref{eq:render})\\ {\scriptsize $p_y(i), p_z(i), g_i, S, G, \vec{e}, \alpha$}};
\node[blk, fill=green!10, right=12mm of fwd] (loss) {Loss (\ref{eq:loss})};
\node[dat, below=6mm of warp] (odo) {Odometry, gyro\\ $t \rightarrow t{+}k$};
\node[dat, below=6mm of loss] (Pf) {Measured\\ $P_t, \ldots, P_{t+T}$};
\draw[arr] (Pt.east) -- (Pt.east -| rb.west);
\draw[arr] (sp.east) -- ++(7.5mm,0) |- (E.west);
\draw[arr] (E) -- (warp);
\draw[arr] (odo) -- (warp);
\draw[arr] (warp) -- node[above, font=\scriptsize]{$\mathcal{W}_k E_t$} (fwd);
\draw[arr] (fwd) -- node[above, font=\scriptsize]{$\hat{P}_{t+k}$} (loss);
\draw[arr] (Pf) -- (loss);
\draw[arr, dashed] (E.north) -- ++(0,5mm) -| node[pos=0.25, above, font=\scriptsize]{$k{=}0$: $\hat{P}_t$ (no warp)} (fwd.north);
\end{tikzpicture}}
\\[1mm]
\includegraphics[width=\linewidth]{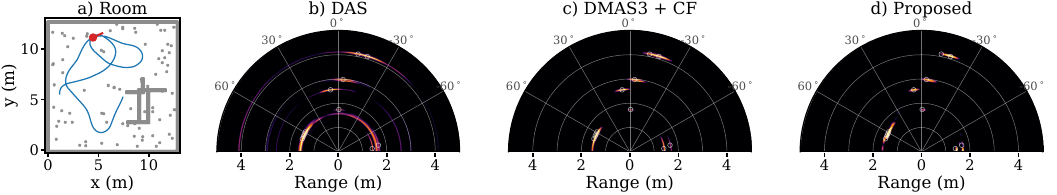}
\caption{Top: overview of the self-supervised training. The network maps the DAS energyscape of pulse $t$ to a reflectivity map, blurred with a fixed kernel $H$ (the output resolution), which is moved to the poses of the next pulses by the acoustic-flow warp (odometry and gyroscope only), rendered by the learned forward model, and compared to the measured energyscapes. Orange: learned; blue: fixed. Bottom: example of one pulse in a validation room (not specifically chosen), as a top view of the energyscape (25~dB dynamic range, $+40\log_{10} r$): a) the room with the path of the robot, b)--d) DAS, third-order DMAS with CF and the proposed method. Circles: strongest true echoes (evaluation only). A video of the validation drives is available at \url{https://youtu.be/N40WfdwaAUs}.}
\label{fig:overview}
\end{figure*}

Figure \ref{fig:overview} (top) shows the model architecture and training scheme. A network $f_\phi$ maps the energyscape $P_t$ to a geometric reflectivity map, which is rendered through equation \ref{eq:render} and compared to $P_t$. More importantly, this map is also warped to the poses of the next $T$ pulses using the acoustic flow equations, and compared to the energyscapes measured there, which will show to be the essential step to learn the deconvolution in a self-supervised manner.

\subsection{Acoustic-flow warp}
For a forward-facing sensor at the center of a platform that moves with linear velocity $V$ and angular velocity $\omega$, a static reflector moves through the 2D image along the acoustic flow \cite{steckelAcousticFlowBasedControl2017, jansenAdaptiveAcousticFlowBased2021}:
\begin{equation}
\frac{dr}{dt} = -V\cos(\theta), \qquad \frac{d\theta}{dt} = \frac{V\sin(\theta)}{r} + \omega
\label{eq:flow}
\end{equation}
We integrate equation \ref{eq:flow} over the motion between pulse $t$ and $t+k$, measured by the wheel odometry ($V$) and the gyroscope ($\omega$), which amounts to a rigid transformation per step. The action-conditioned warp operator $\mathcal{W}_k$ moves every cell of $E_t$ accordingly (backward mapping with linear interpolation) and has no learnable parameters (completely fixed by physics). We warp the internal reflectivity map so that the change of the PSF and of the spreading loss with position is handled by the forward model. Cells that enter the field of view are excluded by a mask $m_k(r,\psi)$.

\subsection{Loss terms}
\textbf{Likelihood:} DAS is a coherent processing operation, so the measured power fluctuates around the incoherent mean of equation \ref{eq:render} as speckle, with an approximately exponential distribution \cite{goodmanFundamentalPropertiesSpeckle1976}. We model $P$ as exponentially distributed around $q = \hat{P} + N_0$, with $N_0$ the noise floor, and make this heavy-tailed with a gamma-distributed precision with $\nu$ degrees of freedom. For $\nu = 1$, the negative log-likelihood per cell, offset to be zero at $q=P$, is given by:
\begin{equation}
\rho(P,q) = 2\log\cosh\left(\frac{1}{2}\log\frac{P}{q}\right)
\label{eq:rho}
\end{equation}
which basically is a squared error in the log domain for small deviations and an absolute error for large ones, so that unexplained power, such as the sidelobes of reflectors entering the field of view, cannot dominate the loss. 

\textbf{Reconstruction and prediction:} With $\hat{P}[E]$ the rendering of equation \ref{eq:render}, the reconstruction and prediction terms are given by:

\begin{align}
\mathcal{L}_\text{rec} &= \frac{1}{MR}\sum_{r,\psi} \rho\left(P_t,\, \hat{P}[E_t] + N_0\right) \label{eq:lrec}\\
\mathcal{L}_\text{pred} &= \frac{1}{T}\sum_{k=1}^{T} \frac{\sum_{r,\psi} m_k\, \rho\left(P_{t+k},\, \hat{P}[\mathcal{W}_k E_t] + N_0\right)}{\sum_{r,\psi} m_k} \label{eq:lpred}
\end{align}
The reconstruction term alone is satisfied by many maps, including the trivial one in which the sidelobes are explained as reflectors. The prediction term breaks this ambiguity: a sidelobe that is explained as a reflector moves along the wrong acoustic flow, and therefore predicts the next energyscapes incorrectly.

\textbf{Sparsity:} The network output $\tilde{E}_t$ is penalized with the concentrating penalty:
\begin{equation}
\mathcal{L}_\text{sp} = \frac{1}{MR}\sum_{r,\psi} \sqrt{\tilde{E}_t(r,\psi) + \epsilon}
\end{equation}
with $\epsilon = 10^{-4}$. The concave square root favors concentrated reflectors, but an $\mathcal{L}_1$ penalty performs equally well (Section \ref{sec:exp}).

\textbf{Sensor priors:} Directions in which reflectors are rarely observed leave the transducer gain $G$ unconstrained. We therefore add a total variation $\mathcal{L}_\text{tv}$ of $G$ (in dB) over neighboring directions, and a squared hinge $\mathcal{L}_\text{ax}$ that keeps $G$ below its on-axis value plus 3~dB. 

\textbf{Final Loss:} The total loss that the network optimizes is the following weighted sum:
\begin{equation}
\mathcal{L} = \mathcal{L}_\text{rec} + \mathcal{L}_\text{pred} + \lambda_\text{sp}\mathcal{L}_\text{sp} + \lambda_\text{tv}\mathcal{L}_\text{tv} + \lambda_\text{ax}\mathcal{L}_\text{ax}
\label{eq:loss}
\end{equation}
with $\lambda_\text{sp}=0.2$, $\lambda_\text{tv}=0.01$ and $\lambda_\text{ax}=0.1$, minimized over the network weights $\phi$ and the parameters $\Theta$ of the forward model jointly.

\subsection{Network and output resolution}
The network takes the compressed energyscape $10\log_{10}(1+P/N_0)$ and the range as input channels, and consists of six blocks with 48 channels (Fig.~\ref{fig:overview}). Each block has three residual branches: a dense mixing over all directions with a learned $M\times M$ matrix (initialized as the identity) \cite{tolstikhinMLPMixerAllMLPArchitecture2021}, as the sidelobes reach the full field of view,  a graph convolution over neighboring directions (Chebyshev filter of order 3 \cite{defferrardConvolutionalNeuralNetworks2016, perraudinDeepSphereEfficientSpherical2019}) and a dilated convolution over range. A softplus yields $\tilde{E}_t \geq 0$ in a smooth manner, and the clean map $E_t = H * \tilde{E}_t$ uses a fixed Gaussian kernel $H$ of 1.5$^\circ$ in azimuth and 0.7 cells in range. This kernel is a design choice as it sets a uniform output resolution over the field of view. The network has 352k parameters and exports directly to ONNX because of the lack of specialized operators.

\section{Experimental Validation}
\label{sec:exp}
\subsection{Simulation setup}
We simulated an eRTIS sensor \cite{laurijssenRuggedizedEmbeddedUltrasound2026} with 32 microphones (Poisson-disc sampling in an ellipse of 7~cm by 5~cm, $f_s = 450$~kHz) and an emitter 4~cm above the array, sending a 3~ms hyperbolic FM sweep from 80~kHz to 20~kHz. The true array deviates from the nominal one by 0.15~mm, 0.8~dB and 5$^\circ$ rms. The emitter has a band-pass response (peak at 45~kHz) and a directivity with a main lobe that narrows with frequency on a diffraction floor at $-13$~dB, as the real transducer falls off much less steeply than a piston model predicts. Its axis is misaligned by 2$^\circ$ in azimuth and 4$^\circ$ in elevation. None of these mismatches are known to the deconvolution model. Echoes were synthesized coherently per microphone, and the energyscapes have $M=181$ directions and 240 range cells of 2~cm. The robot drove smooth paths at 8~cm per pulse through 24 rooms (20 for training, 4 for validation, 500 pulses each) with walls, corners, partitions and point reflectors, giving a realistic 20 to 30 echoes per pulse, with 10\% odometry noise and 0.15$^\circ$ gyroscope noise.

We trained for 15,000 steps with AdamW (learning rate $3\cdot10^{-4}$, 16 windows of $T=4$ future pulses), keeping the forward model fixed for the first 2,000 steps (about 95~min per run). We compare to DMAS, DMAS with CF and third-order DMAS with CF \cite{jansenDelayMultiplyandSumBeamformingRealTime2026}, and to DAMAS \cite{brooksDeconvolutionApproachMapping2006}, Richardson-Lucy (RL, as in DAMAS2 \cite{doughertyExtensionsDAMASBenefits2005}), both with 2,000 iterations, and CLEAN (loop gain 0.2, down to $-40$~dB) on the DAS energyscape. These use the nominal PSF, i.e., equation \ref{eq:render} with the design values, and, as an oracle, the true PSF of the simulator, which corresponds to a perfect calibration. As metrics, we use the fraction of the range-compensated image energy within $\pm 1$ cell and $\pm 3^\circ$ of the 40 strongest true echoes, the contrast between the energy per cell inside and outside these regions, and, for isolated echoes, the $-6$~dB width, the peak error and the detection rate.

\subsection{The learned PSF}
Figure \ref{fig:psf}a) compares the learned PSF $K_\Theta$ to the true PSF, obtained by imaging a point reflector at 2~m with DAS. The learned PSF reproduces the main lobe, its widening towards the sides, and the first sidelobes (dB-correlation 0.95, median error 1.3~dB over all 181 directions). It should be noted that the nominal PSF matches the true shape about as well (0.95, 1.2~dB): in the horizontal plane, the shape is dominated by the known array geometry. What the nominal model cannot know is the sensitivity over the field of view, which falls by 8.2~dB towards 85$^\circ$ due to the emitter directivity. The learned model recovers this curve within 0.5~dB.

\subsection{Image quality}
Figure \ref{fig:psf}b) shows the output of the complete imaging chain for the same reflectors. DMAS suppresses the sidelobes of DAS \cite{jansenDelayMultiplyandSumBeamformingRealTime2026}, but its main lobe still widens from 7$^\circ$ to over 20$^\circ$ towards the sides, whereas the deconvolved output keeps 5$^\circ$ to 6$^\circ$. Fig.~\ref{fig:overview} (bottom) shows an example pulse, and table \ref{tab:results} summarizes the metrics. All methods detect at least 96\% of the isolated echoes, with a median peak error of at most 1$^\circ$. The proposed method concentrates 64\% of the image energy around the true echoes, against 31\% for DAS, 59\% for the best DMAS variant and 56\% for the best classical deconvolution with the nominal PSF, which does not remove the widening towards the sides ($W_{60}$ of 11$^\circ$ to 19$^\circ$). CLEAN with the true PSF reaches 62\%, but requires a calibration, and thousands of iterations instead of one forward pass that can be executed on the highly optimized embedded GPU in the eRTIS sensors.

\begin{table}[t]
\centering
\caption{Image quality (400 validation pulses; mean (std) over three seeds where given). $W_{0}$, $W_{60}$: $-6$~dB width at $|\theta|<15^\circ$ and $60^\circ \le |\theta| < 75^\circ$.}
\label{tab:results}
\setlength{\tabcolsep}{4pt}
\renewcommand{\arraystretch}{0.92}
\footnotesize
\begin{tabular}{lcccc}
\toprule
 & $E_\text{near}\uparrow$ & $C$ (dB)$\uparrow$ & $W_{0}\downarrow$ & $W_{60}\downarrow$ \\
\midrule
DAS & 0.31 & 18.6 & 14$^\circ$ & 33$^\circ$ \\
DMAS & 0.50 & 22.2 & 10$^\circ$ & 24$^\circ$ \\
DMAS+CF & 0.57 & 23.5 & 8$^\circ$ & 21$^\circ$ \\
DMAS3+CF & 0.59 & 23.9 & 8$^\circ$ & 19$^\circ$ \\
\midrule
DAMAS (nom.) & 0.53 & 22.7 & 8$^\circ$ & 19$^\circ$ \\
RL (nom.) & 0.55 & 23.2 & 8$^\circ$ & 18$^\circ$ \\
CLEAN (nom.) & 0.56 & 23.4 & 8$^\circ$ & 11$^\circ$ \\
CLEAN (true) & 0.62 & 24.5 & 7$^\circ$ & 7$^\circ$ \\
\midrule
\textbf{Proposed} & \textbf{0.64} (0.002) & \textbf{24.9} (0.0) & 7$^\circ$ & 7$^\circ$ \\
\midrule
No motion & 0.33 & 18.8 & 12$^\circ$ & 8$^\circ$ \\
Nominal fwd.\ model & 0.59 & 23.9 & 8$^\circ$ & 7$^\circ$ \\
$\mathcal{L}_1$ sparsity & 0.63 & 24.9 & 7$^\circ$ & 7$^\circ$ \\
Log-domain L2 loss & {0.59} (0.006) & {23.9} (0.1) & 6$^\circ$ & 6$^\circ$ \\
Exponential likelihood & 0.51 & 22.4 & 8$^\circ$ & 10$^\circ$ \\
\bottomrule

\end{tabular}
\end{table}

\subsection{Ablations}
The lower part of table \ref{tab:results} lists the ablations. Without the prediction term, the output stays at the level of DAS ($E_\text{near}$ of 0.33). With the forward model fixed at its nominal design, it reaches a score of 0.59, the level of DMAS3 with CF, as the unknown sensitivity is then attributed to the reflectors. A squared error on the compressed levels gives 0.59, and the exponential likelihood 0.51, so the heavy tails of equation \ref{eq:rho} are essential. The $\mathcal{L}_1$ penalty performs on par with the square root.

\textbf{Non-flat emitter spectrum:} With ripples of up to $\pm 10$~dB in the transmit spectrum, the results hardly change: 0.64 when trained on these data, and 0.63 for the model trained on the smooth spectrum.

\section{Conclusion and Future Work}
We presented a self-supervised method to learn the deconvolution of in-air sonar images, together with the PSF of the complete imaging chain, from sonar data and odometry. The motion of the sensor provides the supervision, as true reflectors follow the acoustic flow equations and sidelobes do not. In simulation, with an unknown emitter spectrum and directivity, the deconvolved images reach a constant resolution of 5$^\circ$ to 6$^\circ$ over the full field of view, with 64\% of the image energy around the true echoes, against 59\% for the best DMAS variant and 62\% for CLEAN with the true PSF. As the method requires no ground truth, it can be trained directly on recordings of the real sensor, which is part of our future work. Furthermore, we will publish an extended version of the method that learns 3D point-spread functions. A final step would be to learn the beamformer operation itself from the raw microphone signals.

Claude Opus 5.5 was used for coding, analysis harness generation and writing a first draft. The paper was substantially revised by the authors, and they take full responsability of the content.

\bibliographystyle{IEEEtran}
\bibliography{references}

\end{document}